\documentstyle[a4]{article}
\input epsf
\input diagrams.tex
\input mssymb.sty
\begin{document}
\topmargin=-1cm
\def\ti{\tilde}
\def\Si{\Sigma}
\def\Ga{\Gamma}
\def\ep{\epsilon}
\def\ld{\lambda}
\def\Co{{\Bbb C}}
\def\ha{{\cal H}}
\def\hat{\ti\ha}
\def\hr{\partial}
\def\ra{\rangle}
\def\la{\langle}
\def\NE{{\rm NEFA}}
\def\HI{{\rm HIL}}
\def\ET{{\rm ETA}}
\def\FQ{{\rm FQG}}
\def\kol{\bot}
\def\ss{\subset}
\def\rr{\rightarrow}
\def\sr{\stackrel}
\def\du{\sr{.}{\cup}}      
\def\do{{\cal D}r} 
\def\be{\begin{equation}}
\def\ee{\end{equation}}
\newarrow{TeXto}----{->} 
\begin{center}
{\LARGE Quantum Kramers--Wannier Duality And Its Topology\\}
\vskip 1cm
{\large Pavol \v Severa\\}
\vskip 0.5cm
{\small Dept. of Theor. Physics,
Charles University,\\ \small V Hole\v sovi\v ck\'ach 2, 18000 Prague, Czech
Republic}
\vskip 0.5cm
\end{center}
\begin{abstract}
We show for any oriented surface, possibly with a boundary, how to generalize
Kramers--Wannier duality to the world of quantum groups.
The generalization is motivated by quantization of Poisson--Lie T-duality
from the string theory. Cohomologies with quantum coefficients are defined
for surfaces  and their meaning is revealed. They are functorial with
respect to some glueing operations and connected with q-invariants of
3-folds.
\end{abstract}
\section{Introduction and Summary}
The original motivation for this work was to gain some understanding of 
Poisson--Lie (PL) T-duality \cite{PL}, including zero-modes, at quantum level.
Let us recall the relevant facts: Usual (i.e. abelian) T-duality emerges in the
presence of an abelian group of symmetry and it is a continuous counterpart
of Kramers--Wannier (KW) duality. Its generalization, PL T-duality,
comes from an action of a PL group $G$, which is not necessarily
a symmetry group, but rather gives rise to a conserved charge (`nonabelian
momentum') with values in the dual PL group $\ti G$. In the dual model, the
roles of $G$ and $\ti G$ are interchanged.

There are path integral \cite{akt,tyvon,loop} and renormalization group
\cite{sfet} arguments  for
PL T-duality at quantum level. However, they are far from being complete
and suffer from problems with boundary conditions. In this
respect, it would be nice to have a lattice counterpart of PL T-duality
-- quantum KW duality. We should expect emergence of quantum groups in
quantization of PL T-duality that are deformations of the corresponding
PL groups. It is necessary at least in the case of open
strings \cite{moment}.

The usual (abelian) KW duality is an immediate consequence of
Poincar\'e duality between cohomology groups $H^1(\Si;G)$ and $H^1(\Si;\ti G)$.
Here $\Si$ is a closed oriented surface and $G$, $\ti G$ are mutually dual
finite abelian groups. Surfaces with boundaries are also admissible, provided
one uses appropriate relative cohomologies (to be specified below).

The only problem with quantum KW duality consists in defining a suitable
generalization of $H^1(\Si;G)$, if we replace $G$ with a general (i.e.
non-commutative and non-cocommutative) Hopf algebra, preserving
Poincar\'e duality. In this
paper, this is accomplished for finite quantum groups (finite--dimensional
Hopf $C^*$-algebras). It is easy to define these ``cohomologies'' using
a graph on $\Si$ (via cocycles): ordering problems (coming from
non-commutativity and non-cocommutativity) are absent here, because there
is a natural (cyclic) ordering on the edges around a face (or running from
a vertex). The real problem is to prove topological invariance
of these  ``cohomologies''. We should understand what they really are.

Now I will try to present briefly quantum KW duality (pointing out
the special abelian case) in the form that seems to be the most convenient.

Our basic objects are called neckfaces. These are compact oriented surfaces,
possibly with boundary; on each boundary circle there are some black and
some white points, called beads. The beads are located as on the figure,
i.e. between any two black there is a white one and vice versa;
the beads dissect the boundary into pieces called strings.
$$
\epsfxsize 40mm
\epsfbox{nefa.epsi}
$$
Now suppose
we are given a finite quantum group (a finite-dimensional Hopf $C^*$-algebra)
$\ha$.
Then we will give a construction that provides a vector space $\eta(\Si)$ for
any neckface $\Si$
($\eta(\Si)$ is a non-zero finite-dimensional Hilbert space). It is the
promised analogy of the relative cohomology group $H^1(\Si,B;G)$, where $B$
is the set of black beads (more precisely, it is an analogy of the vector
space $\Co H^1(\Si,B;G)$).
The correspondence $\Si\mapsto\eta(\Si)$ enjoys certain functoriality
properties. We organize neckfaces into a category NEFA: morphisms (glueings)
are maps $f:\Si_1\rr\Si_2$ that may glue pairs of strings of $\Si_1$ (white
endpoints with white and black with black), but up to this they are 
orientation-preserving homeomorphisms. The beads of $\Si_2$ should be
the images
of the beads of $\Si_1$ that lie on $\hr\Si_2$. Now we can see $\eta$
as a functor from NEFA to the category of vector spaces. The properties of
$\eta$ are described in the section 2; for now it is enough to know that
$\eta(\Si_1\du\Si_2)=\eta(\Si_1)\otimes\eta(\Si_2)$.

To have an example in mind, the reader should read the section 3 (and
without inconvenience, she or he may read the exact statements of definitions
given in the section 2). The case $E=1$, $F=G$ will correspond to
$G$-spin models (when $\ha$ is the group algebra $\Co G$); if moreover $G$
is abelian, this will correspond to the usual KW duality. 

The correspondence between $\ha$ and $\eta$ is the following: let $D$ be
a disk with 2 black and 2 white beads. Then $\ha=\eta(D)$. The multiplication
and comultiplication
on $\ha$ are given by two glueings $D\du D\rr D$ that give us (by functoriality)
maps $\ha\otimes\ha\rr\ha$; these are the multiplication and the dual
multiplication (the adjoint of the comultiplication map
$\ha\rr\ha\otimes\ha$, up to a positive factor -- cf. the appendix):
$$
\epsfxsize 50mm
\epsfbox{multip.epsi}
$$
(in the lower diagram, the upper $D$ is the first and the lower $D$
the second in $D\du D$). We use the following convention: images of beads
are drawn as beads even if they are inside and images of glued string are
represented by dotted lines.

Look at the $(G,E,F)$ example: if $E=1$, $F=G$, we clearly obtain what we want:
$(G,E,F)$-bundle type over $D$ is specified by the discontinuity at the white
points, which is an arbitrary element $g\in G$. The first glueing gives
$g_1\otimes g_2\mapsto g_1 g_2$ and the second
$g_1\otimes g_2\mapsto \delta_{g_1,g_2} g_1$; the former is the multiplication
in $\ha=\Co G$ and the latter in the dual algebra $\hat$ of functions on $G$.
In the general case, one obtains the famous Hopf algebra coming from the triple
$(G,E,F)$.

The other operations (unit, counit, $S$ and $*$) are also given by glueings;
we just
mention the antipode $S:\ha\rr\ha$ that comes from rotation of $D$
for 180 degrees
(notice that $S^2=1$ as it should be for finite quantum groups).

There is an important point in the way we recovered the Hopf algebra structure
on $\eta(D)$: we had two maps $\ha\otimes\ha\rr\ha$ and one of them was claimed
to be the multiplication while the other one to be the dual multiplication. The
distinction
was made only by the colour of the beads that were mapped inside $D$.
If we made
it in the other way, we would obtain another Hopf algebra structure
(clearly
the
dual one) on the same space. This identification of $\ha$ with $\hat$ is
given by the Fourier transform and this is the origin of Fourier transforms
in KW duality. In our picture, KW duality will simply state that recolouring
the beads corresponds to replacing $\ha$ with $\hat$. In the case of a
commutative
group $G$ we have $\eta(\Si)=\Co H^1(\Si,B;G)$ ($B$ is the set of black beads);
what we have said implies that this space is naturally identified with
$\Co H^1(\Si,W;\ti G)$ ($W$ is the set of white beads (that becomes $B$ after
recolouring) and $\ti G$ the dual group). The identification
is given by Fourier transform (the groups $H^1(\Si,B;G)$ and
$H^1(\Si,W;\ti G)$ are mutually dual via Poincar\'e duality).

After this preparation we can describe the `statistical models' and their
KW duality. Suppose we are given the surface $nD$ that is simply the disjoint
union of $n$ copies of $D$ and a glueing $f:nD\rr\Si$; for example $\Si$ may
be a square (or a torus) with a tiling by squares coming from $f$.
The model is given by specifying a Boltzmann weight 
$w\in\eta(nD)=\eta(D)^{\otimes n}=\ha^{\otimes n}$.
The `partition sum' is an element of
$\eta(\Si)$: it is simply $\eta(f)(w)$. If it is necessary, we may take
its inner product with some $c\in\eta(\Si)$ to obtain a number;
one may say that $c$ specifies the boundary and periodicity conditions.

The correspondence with the usual lattice models is the following:
we have the surface $\Si$ tiled by quadrilaterals (the tiling comes from $f$);
every vertex of the tiling has a colour (given by its proimage in $nD$ which is
a set of beads of the same colour). We can construct two graphs on $\Si$:
draw the diameter connecting the black beads on each $D$ in $nD$; their
images form a graph $\Ga\ss\Si$.
If we make the same thing with the white beads,
we obtain a graph $\ti\Ga$, dual to $\Ga$. One can go also in the
opposite direction, starting with $\Ga$, making $\ti\Ga$ and finally
arriving at
the tiling $f$. In the usual formulation of KW duality, there are $G$-valued
spins at vertices of $\Ga$ interacting through the edges; the dual model lives
on $\ti\Ga$ and has $\ti G$-valued spins (where $\ti G$ is the dual of 
the finite abelian group $G$). Only internal spins are summed over (external
spins form the boundary condition) and the spins have to be allowed to take
values in an arbitrary principal $G$-bundle over $\Si$. The partition sum
depends on these boundary and periodicity conditions. In the picture
described above, we organized these sums to a single element of a vector space.

Recall that in our picture KW duality simply states
that recolouring corresponds to transition to the dual Hopf algebra and that
on $D$ the recolouring gives Fourier transform (recoloured $D$ is after 
a rotation again $D$). On $nD$ one has to make Fourier for each $D$.
$$
\epsfxsize 80mm
\epsfbox{graph.epsi}
$$

There remain several things that should be added:

We presented the functor $\eta$ as coming from a finite quantum group $\ha$
and showed how to reconstruct $\ha$ from $\eta$. The functor $\eta$ was
motivated as a generalization of cohomology to quantum coefficients.
But our final formulation,
given in the section 2, characterizes $\eta$ by a set of axioms (not
involving Hopf algebras) and then the theorem, encompassing all our
results, claims that there is a 1-to-1 (up to isomorphisms) correspondence
between such functors $\eta$ and finite quantum groups (in fact it claims
a little bit more -- that the two categories are equivalent).

In the PL T-duality, an important role is played by the Drinfeld double
of the PL groups. It should be expected in the KW duality as well; in fact,
in the case of $G$-spin models (with possibly nonabelian $G$) it was found
that the order and disorder operators are organized into the Drinfeld double
and also that the canonical $R$-matrix of the double gives the braid group
statistics of such models \cite{szve}. The same is true generally and
becomes self-evident when the double is realized as $\eta(\Si)$ for
certain $\Si$ (an anulus with one black and one white bead on each
boundary circle). This will be done in the section 5.

Finally, we can naturally generalize `statistical' models. They were
described by a glueing $f:nD\rr\Si$ and a weight in $\eta(nD)$. Of
course, it is possible to take an arbitrary glueing $f:\Si'\rr\Si$ and
$w\in\eta(\Si')$. If $\Si'$ is a disjoint union of disks (but this
time with an arbitrary number of beads), we obtain duality for 
many-particle interactions (this interpretation is for $\ha=\Co G$).
Just imagine (as we did in the case of $nD$) $\Si$ tiled by polygons
with coloured vertices. At each black vertex there is a $G$-valued
spin and at each polygon an interaction. This may be nice in
renormalization, which consits just in writing the map $f:nD\rr\Si$
as a composition $nD\rr\Si'\rr\Si$.

\section{The Theorem}

A {\em neckface} is a compact oriented surface $\Si$, possibly with a boundary,
together with two finite subsets $B$ ({\em black beads}) and $W$ ({\em white beads})
of $\hr\Si$, such that
\begin{itemize}
\item{
$B\cap W=\emptyset$}
\item{
for each boundary circle $c$, $B\cap c\ne\emptyset$, $W\cap c\ne\emptyset$ }
\item{between any two black beads there is a white one and vice versa.}
\end{itemize}
The part of $\hr \Si$ between two neighbouring beads is called {\em a string}.
\\

The category $\NE$ has neckfaces as its objects; the morphisms ({\em glueings})
are continuous maps $f:\Si_1\rr\Si_2$ such that
\begin{itemize}
\item{
$f$ is onto}
\item{
$f|_{{\rm int}\Si_1}$ is an orientation-preserving
homeomorphism onto its image}
\item{
for any string $s\ss\Si_1$, either $f$ maps $s$ bijectively onto a string
$s_2\ss\Si_2$, preserving the colour of the endpoints, or there is
another string
$s'\ss\Si_1$ that gets glued with $s$ 
(white end with white and black with black)}
\end{itemize}
Clearly, $f^{-1}(B_2)\ss B_1$, $f^{-1}(W_2)\ss W_1$; if $f^{-1}(B_2)= B_1$ and
$f^{-1}(W_2)= W_1$, i.e. if no bead gets lost inside $\Si_2$, the glueing $f$ is
called {\em nice}.
\\

The object of category $\HI$ are non-zero finite-dimensional Hilbert spaces;
morphisms are linear maps. A morphism $f:H_1\rr H_2$ is {\em projecting},
if the restriction $(\ker f)^{\kol}\rr H_2$ is unitary.
\\

A functor $\eta:\NE\rr\HI$ is an {\em $\eta$-functor} if
\begin{enumerate}
\item{
$\eta(\overline\Si)=\overline{\eta(\Si)}$ ($\overline\Si$ is $\Si$ with
the opposite orientation) }
\item{
$\eta(\Si_1\du\Si_2)=\eta(\Si_1)\otimes\eta(\Si_2)$ }
\item{
for any glueing $f$, $\eta(f)$ is projecting }
\item{
if $f$ is a nice glueing, $\eta(f)$ is invertible (and thus unitary) }
\item{
({\em independent projection law}) Let
$$
\begin{diagram}[small]
&&\Si_1&&\\
&\ldTo<{f_2}&&\rdTo>{f_3}&\\
\Si_2&&&&\Si_3\\
&\rdTo<{g_2}&&\ldTo>{g_3}&\\
&&\Si_4&&\\
\end{diagram}
$$
be a commutative square and let
$f=g_2\circ f_2=g_3\circ f_3$.
Then in $\eta(\Si_1)$ the orthogonal projections onto
$(\ker\eta(f_2))^\kol$, $(\ker\eta(f_3))^\kol$ commute.
If moreover the only beads of
$\Si_1$ that get lost inside $\Si_4$ are those that get lost either inside
$\Si_2$ or in $\Si_3$ then 
$$(\ker\eta(f_2))^\kol\cap(\ker\eta(f_3))^\kol=(\ker\eta(f))^\kol $$
(such squares will be called {\em minimal}). }
\end{enumerate}

{\bf Remark:}
The condition 4 could be avoided by adding to $\NE$ new morphisms
generated by the old ones and inverses to nice glueings.
These are no longer maps
(they involve cutting). We will not do this because of Haupt\-vermutung-like
problems, but heuristically it is important. At any rate, 4 is a special case
of 5.\\
Because $\eta(f)$ is projecting for any $f:\Si_1\rr\Si_2$, we may see
$\eta(\Si_2)$ inside $\eta(\Si_1)$ as $(\ker\eta(f))^\kol$ and $f$ as the
orthogonal projection. The condition 5 states that this is consistent inside
commutative squares; if moreover the square is minimal, the total projection
(corresponding to $f$ in condition 5) is simply the composition of the 
projections
corresponding to $f_2$ and $f_3$. \\

Let $D$ be a disk with two black and two white beads; for any $\eta$-functor
one can find a finite quantum group structure on the space $\eta(D)$ (cf. the
appendix for the definition of finite quantum groups and their category FQG);
this structure comes from glueing and will be described below. A part of our
theorem claims that this gives us a bijective correspondence between
$\eta$-functors and finite quantum groups, up to isomorphisms.

For the full statement of the theorem, we have to organize $\eta$-functors
into {\em category $\ET$}. A morphism $\eta_1\rr\eta_2$ is a collection
of linear maps $\eta_1(\Si)\rr\eta_2(\Si)$ preserving the inner products
(i.e. isometric injections) such that the diagram
\begin{diagram}[heigth=2em,w=4em]
\eta_1(\Si) & \rTo^{\eta_1(g)} & \eta_1(\Si') \\
\dTo        &                  & \dTo         \\
\eta_2(\Si) & \rTo^{\eta_2(g)} & \eta_2(\Si') \\
\end{diagram}
commutes {\em up to a positive factor} for each glueing $g:\Si\rr\Si'$.
Now the map $\ET\rr\FQ$, $\eta\mapsto\eta(D)$ becomes a functor.
\\

{\bf Theorem:}{\it\ 
 The functor $\ET\rr\FQ$, $\eta\mapsto\eta(D)$ is an equivalence of
categories.}
\\

We have to add one thing to the statement of the theorem. 
For any neckface $\Si$, let $\Si^{ex}$ denote $\Si$ with exchanged colours
of the beads. For any $\eta$-functor $\eta$ there is another $\eta$-functor
$\ti\eta$
given by $\ti\eta(\Si)=\eta(\Si^{ex})$. Clearly, the FQG corresponding
to $\ti\eta$ is $\ti\ha$. {\em This fact}
(that the automorphism $\eta\mapsto\ti\eta$
is translated to the duality automorphism){\em is Kramers--Wannier duality}.

Let us look at the role of the Fourier transform in KW duality. Clearly,
$D^{ex}$ may be identified with $D$ after a rotation, i.e we have a map
$\ha\rr\ti\ha$. It is the Fourier transform: Consider the
reflection of $D$ with respect to the north-west--south-east diameter.
It is a map $D\rr\overline{D^{ex}}$. Using $\eta$, it gives us
 $\eta(D)\rr\overline{\eta(D^{ex})}$, i.e. $\ha\rr\overline{\ti\ha}$.
It exchanges multiplication with the dual one, $S$ with $S\circ *$, etc.
So it is nothing else but the trivial map coming from the fact that
$\ti\ha=\overline\ha$ (cf. Appendix). Now compose this reflection with the
reflection with respect to the vertical diameter (the one giving $*$ when
mapped by $\eta$). The result is the mentioned rotation, and when mapped
by $\eta$, it gives us the Fourier transform $\ha\rr\ti\ha$.

\section{An Example}

Let $G$ be a finite group and $E$, $F$ its subgroups such that every element
$g\in G$ can be written uniquely as a product $g=ef$, $e\in E$, $f\in F$.
By a $(G,E,F)$-bundle over a neckface $\Si$ we mean a principal $G$-bundle
over $\Si$ with a specified lift of each string to the bundle; discontinuities
at the beads are allowed but at the black beads they have to be from $E$
and at the white beads from $F$.

An automorphism of a $(G,E,F)$-bundle is an automorphism of the $G$-bundle
preserving the liftings of the strings. Clearly, nontrivial automorphisms
can exist only on the closed components of $\Si$.

Let us fix a triple $(G,E,F)$.
Let $X(\Si)$ be the set of all $(G,E,F)$-bundle types over $\Si$. We put
$$ \eta(\Si)=\Co X(\Si) $$
(the free vector space over $X(\Si)$).
The inner product on $\eta(\Si)$ is defined as follows: the basis $X(\Si)$ is
orthogonal and for $x\in X(\Si)$, $$\la x,x\ra=\mbox{the number of automorphisms
of }x. $$

If $f:\Si_1\rr\Si_2$ is a glueing,
we define $\eta(f):\eta(\Si_1)\rr\eta(\Si_2)$
as follows: for $x\in X(\Si_1)$ we try
to extend the glueing $f$ to the bundle in
such a way that if two strings are glued,
we glue their lifts. If it is possible,
the result is $\eta(f)(x)\in X(\Si_2)$; otherwise $\eta(f)(x)=0$.

This functor is an $\eta$-functor, up to one problem: the maps $\eta(f)$ are
projecting only after rescaling the inner products. But if we rescale each
$\eta(f)$ by a positive number so as to obtain a projecting map, we obtain
a genuine $\eta$-functor.

\section{The proof}

The strategy of the proof is as follows:

First we define the finite quantum group  structure on $\ha=\eta(D)$ for any
$\eta$ (this definition
should be cosidered as a part of the statement of the theorem)
and verify that it obeys FQG axioms.

Then we prove that $\eta$ is specified by the FQG $\ha=\eta(D)$ up to
isomorphisms. The point is that any neckface can be glued from several copies
of $D$. Representations of some basic glueings by $\eta$ are already present
in $\ha$ (by its definition); this is sufficient to restore $\eta$.

Finally, we have to show that there is an $\eta$ for any $\ha$. Our original
motivation is used here -- $\eta$ is constructed as a generalization of
cohomologies to quantum coefficients.

In this proof we show explicitely only that $\eta\mapsto\eta(D)$ is a bijection
from ETA to FQG, up to isomorphisms. In Section 4.3 we construct a functor
$\FQ\rr\ET$. The fact that the composition $\ET\rr\FQ\rr\ET$ is isomorphic
to the identity follows from the proof immediately. 

\subsection{$\ha=\eta(D)$ is a finite quantum group}

First notice the following figure:
$$
\epsfxsize 50mm \epsfbox{dzero.epsi}
$$
It represents a {\em nice} glueing $C\du C\rr C$, so we have an {\em
isomorphism} $\eta(C)\otimes\eta(C)\rr\eta(C)$. As a consequence,
$\eta(C)=\Co$. Also notice that $\eta(S^2)$ is 1-dimensional, because
there is an obvious glueing $C\rr S^2$.

Now we can define the FQG structure on $\eta(D)$. The multiplication
and the dual multiplication were already given on the figure 
at page 2. The counit
is given by the following picture:
$$
\epsfxsize 50mm \epsfbox{epsil.epsi}
$$
It represents
two maps $D\rr C$; the bottom object is a sphere and I indicated on
$S^2$ the image of $\hr D$ (rather than $\hr C$) to make clear that
the diagram is commutative. According to the independent projection
law, the two glueings are represented by the same
$\ep:\eta(D)\rr\eta(C)=\Co$. The definition of $\ti\ep$ is similar (just
change the colour of the beads).

Finally, the involution $*:\eta(D)\rr\overline{\eta(D)}$ is given by
the reflection with respect to the diameter connecting the white beads
of $D$ and the antipode $S$ by rotating $D$ for 180 degrees.

Now we have to prove that we really defined a finite quantum
group. The only problem is to show that the comultiplication acts
as a homomorphism with respect to the multiplication. The next picture
presents another definition of comultiplication, where this condition
becomes evident:
$$
\epsfxsize 50mm \epsfbox{comult.epsi}
$$
It has the form $D\du\Si\rr\Si$ and a nice glueing $D\du D\rr\Si$ is
indicated to show that $\eta(\Si)=\ha\otimes\ha$. We claim that the
corresponding map $\ha\otimes(\ha\otimes\ha)\rr\ha$ is given by 
$a\otimes(b\otimes c)\mapsto a_{(1)}b\otimes
a_{(2)}c$.
In other words, we have to prove something about the following diagram
(the dashed arrows indicate nice morphisms):
$$
\epsfxsize\hsize \epsfbox{diag1.epsi}
$$
Consider the image of the diagram by the functor $\eta$. There are two
paths from the leftmost object to the rightmost one; if we have to go
against an arrow, we take the adjoint of the map. We have to prove
that the lower path is just $\la\ep,\ti\ep\ra$ times the upper path.
So, consider the following diagram:
$$
\epsfxsize 70mm \epsfbox{diag2.epsi}
$$
It was obtained from the previous one by nice glueings (the dashed arrows
are now contracted to identities). It has the same representation by
$\eta$, so that we have to prove the same thing about the new
diagram. And it follows from the following general fact:
$$
\epsfxsize 60mm \epsfbox{lemma.epsi} 
$$
On this diagram, two glueings of the form $\Si'\rr\Si$ are drawn (it
represents only the relevant parts of general neckfaces); also a nice
glueing $\Si\du D\rr\Si'$ is indicated. We claim that having mapped the
diagram to $\HI$ by
$\eta$, going from the left to the right is just $\la\ep,\ti\ep\ra$
times the identity. And the nice glueing $\Si\du D\rr\Si'$ makes it
clear.

\subsection{$\eta$ is determined uniquely by $\ha=\eta(D)$}

If $\Si$ is a neckface then the space $\eta(\Si)$ becomes an
$\ha$-module for every black bead of $\Si$ and an $\hat$-module for
every white bead. The picture represents the black bead case:
$$
\epsfxsize 50mm \epsfbox{hmod.epsi}
$$
Now suppose we are given a glueing $f:\Si'\rr\Si$. These $\ha$ and $\hat$
module structures on $\eta(\Si')$ will enable us to write down the
projector $\eta(\Si')\rr(\ker\eta(f))^\kol$. The simplest case is the
following:
$$
\epsfxsize 30mm \epsfbox{glu0.epsi}
$$
Here the projector is given obviously by the action of $\ep\in\ha$
at the black bead (due to the definition of $\ep$). The recoloured
version is similar.

Consider the following neckface (with a nice glueing $\Si'\rr\Si$
indicated):
$$
\epsfxsize 15mm \epsfbox{hten.epsi}
$$
In $\Si'$ the black bead splits into 3 beads (3 is just an example);
they are numbered against
the orientation of $\Si$. By the nice glueing we have $\eta(\Si')\simeq
\eta(\Si)$. Upon this identification, the action of $a\in\ha$ at
the black bead of $\Si$ is the same as the action of $a_{(1)}\otimes
a_{(2)}\otimes a_{(3)}$ at the 3 black beads of $\Si'$. This follows
immediately from the bottom figure on page 7. A similar thing is valid for
any number and in the recoloured version.

Putting these two facts together, we can say something about the
following
type of glueing $\Si'\rr\Si$:
$$
\epsfxsize 30mm \epsfbox{gpro.epsi}
$$
The projector $\eta(\Si')\rr(\ker\eta(f))^\kol$ is simply the action
of $\ep_{(1)}\otimes
\ep_{(2)}\otimes \ep_{(3)}$ at the 3 black beads of $\Si'$.

Now we can state the general form of the projector
$\eta(\Si')\rr(\ker\eta(f))^\kol$. If $n$ black beads of $\Si'$ are
mapped to a single point {\em inside} $\Si$, we have to act by
$\ep_{(1)}\otimes\ep_{(2)}\dots\otimes \ep_{(n)}$ (as in the previous
figure). We call this operator {\em a black projector}.
Similar {\em white projection} has to be made at each image of white
beads inside $\Si$. These projections are mutually commuting and their
composition is the projector $\eta(\Si')\rr(\ker\eta(f))^\kol$. This
can be proved from the fact concerning the previous picture and from the
independent projection law.

Now we can prove that the finite quantum group structure on
$\ha=\eta(D)$
determines $\eta$ uniquely. The point is that any neckface $\Si$ can
be glued from several copies of $D$ and the $\ha$ and $\hat$ module
structure on $\eta(D)$ is given simply by the multiplication and the
dual multiplication. Thus we know $\eta(\Si)$ as a subspace of
$\ha^{\otimes n}$ and we also know $\eta(f)$ for any
$f:\Si_1\rr\Si_2$.

\subsection{Construction of $\eta$ for a given $\ha=\eta(D)$:\\
Cohomologies with quantum coefficients}

Now we know much about reconstructing $\eta$ from its $\ha$.
For any neckface $\Si_1$ we take a glueing $f:nD\rr\Si_1$ and we
already know the subspace
$\eta(\Si_1)\simeq(\ker\eta(f))^\kol\ss\ha^{\otimes n}$.
Also, any glueing $g:\Si_1\rr\Si_2$ gives us a glueing $nD\rr\Si_2$ (by
composition), so we know $\eta(\Si_2)$ and also $\eta(g)$ (in this
picture, $\eta(\Si_2)\ss\eta(\Si_1)$ and $\eta(g)$ is the orthogonal
projection). The only thing we miss is the identification of
$\eta(\Si)$ for any two glueings $nD\rr\Si$ and $mD\rr\Si$: we see
$\eta(\Si)$ as a subspace of $\eta(nD)$ and also of $\eta(mD)$; we have
to understand how to identify these subspaces, using the finite
quantum group $\ha$ only. Clearly, this has to be possible (in the
view of the previous section).

Now we simply reverse the line of thoughts. Suppose we are given $\ha$ and we
want to define a corresponding $\eta$ (we already know that $\eta$ is unique
up to an isomorphism). The method is rather straightforward: for any glueing
$f:nD\rr \Si$ we define $\eta_f(\Si)$ as a subspace of $\ha^{\otimes n}$ in the
way described above -- as the common range of the projectors like
$\ep_{(1)}\otimes\ep_{(2)}\dots\otimes \ep_{(n)}$. One easily verifies that
these projectors are orthogonal and mutually commuting. For any glueing
$g:\Si\rr\Si'$ we have $\eta_{g\circ f}(\Si')\ss\eta_f(\Si)$. We define
$\eta_f(g):\eta_f(\Si)\rr\eta_{g\circ f}(\Si')$ as the orthogonal projection.
To define $\eta$ itself we have to show unitary isomorphisms
$\Xi_{f_2,f_1}:\eta_{f_1}(\Si)\rr \eta_{f_2}(\Si)$ such that
\be \Xi_{f_3,f_2}\circ\Xi_{f_2,f_1}=\Xi_{f_3,f_1}.\ee
Then we may identify
$\eta_f(\Si)$'s for any two $f$'s and thus define $\eta(\Si)$.
If $g:\Si\rr\Si'$ is a glueing, we define $\eta(g)$ as $\eta_f(g)$
for any $f$.  Of course we have to require
\be 
\begin{diagram}
\eta_{g\circ f_1}(\Si')               & \ss & \eta_{f_1}(\Si)       \\
\dTo^{\Xi_{g\circ f_2,g\circ f_1}}    &     & \dTo_{\;\Xi_{f_2,f_1}}\\
\eta_{g\circ f_2}(\Si')               & \ss & \eta_{f_2}(\Si)       \\
\end{diagram}
\ee
to be commutative for any $f_{1,2}:n_{1,2}D\rr\Si$.
If (1) and (2) hold, we have a well
defined functor $\eta$. Notice that it automatically satisfies all
the axioms of an $\eta$-functor, except possibly for
$\eta(\overline\Si)=\overline{\eta(\Si)}$.
If we can check this and the fact that the
FQG derived from $\eta$
is $\ha$, we are done.

All the rest of this section is devoted to a description of an appropriate
$\Xi$.

We use the graphs corresponding to a glueing $f:nD\rr\Si$, such as on the
figure on page 3. The graphs will be called $\Ga_f$ (black vertices) and
$\ti\Ga_f$ (white vertices), or the black graph and the white graph. Let $E_f$
denote the set of edges of $\Ga_f$ and $\ti E_f$ the set of edges of 
$\ti\Ga_f$.
We choose an orientation of each edge of
$\Ga_f$ and $\ti\Ga_f$ in the following way: on each $D$ we orientate
the horizontal diameter (connecting the black beads)
 from the left to the right and
the vertical diameter from the top to the bottom; the edges of $\Ga_f$   
and $\ti\Ga_f$ are their images. There is a 1-to-1 correspondence
between the edges of $\Ga_f$ (and also $\ti\Ga_f$) and the disks
among $nD$.

We will consider intersection of graphs (always a black graph with a
white graph, but not necesarilly coming from the same glueing
$nD\rr\Si$). We will always suppose that the two graphs have only
finitely many common points, none of them being a vertex, and that the edges
intersect transversally. This can be achieved by a deformation.
An intersection point is {\em positive} or {\em negative} according to the
relative orientation of the intersecting edges;
by definition, all the intesections
of $\Ga_f$ with $\ti\Ga_f$ are positive.

We will consider intersections of a
$\Ga$ with graphs in a 1-parametric continuous family $\ti\Ga_\ld$,
$\ld\in\la0,1\ra$, too. In this case we admit only finitely many $\ld$'s for
which the intersection of $\Ga$ and $\ti\Ga_\ld$ violate the previous
condition, always in this way: either an edge of $\ti\Ga$ is crossing
a vertex of $\Ga$, or a vertex is crossing an edge, or two opposite
crossing of edges disappear or appear; only one such exception is
allowed for a fixed $\ld$:

$$
\epsfxsize 45mm \epsfbox{croscond.epsi}
$$
This situation can be achieved by a small deformation.

Before describing the system $\Xi$ itself, we describe a similar procedure
for a combinatorial definition of $H^1(\Si,B;G)$. Recall (Section 3)
 that $\eta(\Si)=\Co H^1(\Si,B;G)$ if $\ha=\Co G$
for a finite abelian group $G$ ($B$ is the set of the black beads of $\Si$).
Changing the procedure slightly we arrive at a combinatorial definition of
$\Co H^1(\Si,B;G)$ using the Hopf algebra structure of $\Co G$ only.
This is easily generalized to an arbitrary FQG (just by taking care
of ordering problems). This generalization will be our definition of $\Xi$.

One can define $H^1(\Si,B;G)$ combinatorially
as cocycles on the graph $\Ga_f$ modulo
coboundaries, i.e. modulo action of $G$ at the internal vertices of $\Ga_f$.
Here the transfer of cocycles on $\Ga_{f_1}$ to 
 $\Ga_{f_2}$ can be done in the following way: take the graph $\ti\Ga_{f_1}$;
if $c_1$ is a cocycle on $\Ga_{f_1}$, the corresponding
 cocycle $c_2$ on $\Ga_{f_2}$
 asigns to an edge $e_2$ the product $\prod c_1(e_1)^{\pm 1}$.
Here the product runs over
all the intersections of $e_2$ with the edges of $\ti\Ga_{f_1}$. If it is the
intersection with an edge $\ti e_1$, we put to the product $c_1(e_1)^{\pm 1}$
according the sign of the intersection
(here $e_1\in E_{f_1}$ is the edge dual to $\ti e_1$, i.e. corresponding
to the same $D$ in $n_1D$).

Now we come to a combinatorial definition of $\Co H^1(\Si,B;G)$.
Let $f:nD\rr\Si$ be a glueing. Any element of  $H^1(\Si,B;G)$ is represented
by a cocycle on the graph $\Ga_f$, i.e. by an element of $G^n$. Of course,
this cocycle is not unique: we can act by an element of $G$ at any internal
vertex of $\Ga_f$. But suppose we take $(\Co G)^{\otimes n}=\Co (G^n)$
instead of $G^n$. In this space we can average over the action of $G$;
the averaging is nothing but the black projector.
So we get an element of $(\Co G)^{\otimes n}$ which is now unique. It is
{\em closed} (i.e. it is a linear combination of cocycles)
and {\em coclosed} (i.e. $G$-invariant at the internal vertices of $\Ga_f$);
therefore we may call it {\em harmonic}. By linearity, each element of
$\Co H^1(\Si,B;G)$ is represented by a unique harmonic element of
$(\Co G)^{\otimes n}$; in fact, $\Co H^1(\Si,B;G)$ may by identified
with the space of harmonic elements of $(\Co G)^{\otimes n}$. Evidently,
closedness is equivalent to invariance with respect to the white projectors,
i.e. the space of harmonic elements of $(\Co G)^{\otimes n}$ is just
 $\eta_f(\Si)$. Generally, elements of the range of the black (white)
projectors will be
called coclosed (closed). The procedure given for $ H^1(\Si,B;G)$
works for  $\Co H^1(\Si,B;G)$ almost without any change: we just have to
use comultiplication to make tranformations like
$g\mapsto g\otimes\dots\otimes g$ linear. The procedure has to be concluded
by the black projectors (i.e. averaging) to assure coclosedness. The details
will be given below for a general $\ha$.

Recall the definition of the black projectors.
Let $P$ be a vertex of $\Ga_f$ and let 
$e_1,e_2,\dots,e_k$ be the edges incident with $P$, in the cyclical
order, given by the orientation of $\Si$. Some edges may occur
twice in this list. Let
$$a=\bigotimes_{e\in E_f}a_e\in\ha^{\otimes E_f}=\ha^{\otimes n}.$$
If all the edges point to $P$, we form the
coproduct $\ep_{(1)}\otimes\dots\otimes\ep_{(k)}$ and multiply each
$a_{e_i}$ by $\ep_{(i)}$ from the right. If
we allow arbitrary orientations of $e_i$'s, we multiply by
$S(\ep_{(i)})$ from the left instead, if $e_i$ is pointing from $P$.

Let $b\in\ha$. Let us replace $\ep$ with $b$ in the definition of
the black projector. Clearly, $a$ is coclosed at $P$ iff the result
is $\la\ep,b\ra a$ for any $b$. A similar fact is of course valid for 
closedness (just using $\ti\ha$ instead of $\ha$). It can be translated
to the following formulation, very similar to closedness of a group-valued
cochain.

Let $F$ be an internal face of $\Ga_f$ (i.e. containing an internal vertex
of $\ti\Ga_f$). 
Let $e_1,e_2,\dots,e_k$ be
the edges of $\Ga_f$ forming the boundary $\hr F$ of the face $F$, in the order
given by the orientation of the face.
We make comultiplication on $a=\otimes a_e$ at the edges
among $\hr F$ (it may happen that an edge is twice in $\hr F$; in that
case we make the comultiplication twice). For the sake of clarity,
first suppose that all the edges $e_1,e_2,\dots,e_k$ have the
orientation compatible with $F$. In that case, we take the first
component of each coproduct and multiply them in the order
$1,\dots,k$; we have something like
$$a_{e_1(1)}a_{e_2(1)}\dots a_{e_k(1)}\otimes a_{e_1(2)}\otimes a_{e_2(2)}
\otimes\dots\otimes a_{e_k(2)}
\bigotimes_{e\notin\{e_i\}}a_e\in\ha\otimes\ha^{\otimes E_f}.$$
The
condition states that this has to be $1\otimes a$. If the
edges $e_1,e_2,\dots,e_k$ have arbitrary orientation, we also make the
comultiplication on each $a_{e_i}$, but if the orientation of $e_i$ is
opposite to $F$, we take from the coproduct $a_{e_i(2)}$ and put to
the product $S(a_{e_i(2)})$; again we have to obtain $1\otimes a$.
If an edge $e_i$ is twice in $\hr F$, we take to the product
$a_{e_i(1)}$ and $S(a_{e_i(3)})$ (the orientation of $e_i$ once agrees
and once disagrees with $F$). The condition $a\mapsto1\otimes a$ is clearly
independent of the choice of $e_1$.

We use this definition of closedness in what follows; it simplifies the
proof.

Now we define the system of unitary isomorphisms
$\Xi_{f_2,f_1}:\eta_{f_1}(\Si)\rr\eta_{f_2}(\Si)$. Let
$I_{f_2,f_1}=\Ga_{f_2}\cap\ti\Ga_{f_1}$. Recall our requirements on the
intersection of the graphs $\Ga_{f_2}$ and $\ti\Ga_{f_1}$; if necessary,
we deform $\ti\Ga_{f_1}$ so that they held. Later we shall prove that
$\Xi_{f_2,f_1}$ is independent of deformations of the graphs.

We define $\Xi_{f_2,f_1}$ as the composition of several maps. First, let
us define a map $\ha^{\otimes\ti E_{f_1}}\rr\ha^{\otimes I_{f_2,f_1}}$ 
(recall $\eta_{f_1}(\Si)\ss\ha^{\otimes E_{f_1}}=\ha^{\otimes\ti E_{f_1}}$).
For any edge $\ti e\in\ti E_{f_1}$ we take the points
$p_1^{\ti e},\dots,p_{k_{\ti e}}^{\ti e}\in I_{f_2,f_1}$
lying on $\ti e$, in the natural
order given by the orientation of $\ti e$. Given an element
$$\bigotimes_{\ti e\in\ti E_{f_1}}a_{\ti e}\in\ha^{\otimes\ti E_{f_1}}$$
we form the coproduct
$$\bigotimes_{\ti e\in\ti E_{f_1}}(a_{\ti e(1)}\otimes\dots\otimes
a_{\ti e(k_{\ti e})})\in\ha^{\otimes I_{f_2,f_1}}$$
(if $k_{\ti e}=0$, we act by the counit on $a_{\ti e}$);
$a_{\ti e(m)}$ corresponds to $p_m^{\ti e}$.

Now we define a map
$\ha^{\otimes I_{f_2,f_1}}\rr\ha^{\otimes I_{f_2,f_1}}$.
We simply act by the antipode at the negative points of $I_{f_2,f_1}$.

The next map is $\ha^{\otimes I_{f_2,f_1}}\rr\ha^{\otimes E_{f_2}}$.
For any $e\in E_{f_2}$ take its points $p_1^e,\dots,p_{l_e}^e\in I_{f_2,f_1}$
in the natural order. For
$$\bigotimes_{p\in I_{f_2,f_1}}a_p\in\ha^{\otimes I_{f_2,f_1}}$$
form
$$\bigotimes_{e\in E_{f_2}}a_{p_1^e}\dots a_{p_{l_e}^e}.$$

The final map $\ha^{\otimes E_{f_2}}\rr\ha^{\otimes E_{f_2}}$
ensures coclosedness: it is just the composition of all the black
projectors of $f_2$.

$\Xi_{f_2,f_1}$ is equal to the composition
$$
\eta_{f_1}(\Si)\ss\ha^{\otimes\ti E_{f_1}}\rr\ha^{\otimes I_{f_2,f_1}}
\rr\ha^{\otimes I_{f_2,f_1}}\rr\ha^{\otimes E_{f_2}}\rr\ha^{\otimes E_{f_2}}
$$
times a positive factor $c_{f_2,f_1}$ ensuring unitarity. One easily sees
that $\mbox{rng }\Xi_{f_2,f_1}\ss\eta_{f_2}(\Si)$: we only have to check
the evident fact that the composition
$$\ha^{\otimes E_{f_1}}=\ha^{\otimes\ti E_{f_1}}\rr\ha^{\otimes I_{f_2,f_1}}
\rr\ha^{\otimes I_{f_2,f_1}}\rr\ha^{\otimes E_{f_2}}$$
preserves closedness. The factor $c_{f_2,f_1}$ is
$$c_{f_2,f_1}=\la\ep,\ti\ep\ra^{\ti v_{f_2}-\ti v_{f_1}\over 2}$$
where $\ti v_f$ is the number of vertices of $\ti\Ga_f$.

First we have to prove that $\Xi_{f_2,f_1}$ is independent of deformations
of the graphs; this is a straightforward consequence of the useful identity
$\ep_{(1)}a\otimes\ep_{(2)}=\ep_{(1)}\otimes\ep_{(2)}S(a)$.
Next we have to check (1). We do it in the case when $f_2$ and $f_3$ are
{\em close} to each other (in the sense specified below) and then use the
fact that any $f_2$, $f_3$ can be connected by a sequence
$f_2=f^1,\dots,f^n=f_3$ with close $f^i$ and $f^{i+1}$.

We say that glueings $f_{2,3}:n_{2,3}D\rr\Si$ are close to each other if
the graphs $\Ga_{f_{2,3}}$ differ just by removing an edge separating two
different faces or by contracting an edge connecting two different
vertices. For close $f_{2,3}$ the equation (1) is straightforward; we
just have to draw the graphs (in the case of contraction)
so that the contracted graph follows closely the original one, as indicated
on the figure:

$$
\epsfxsize 70mm
\epsfbox{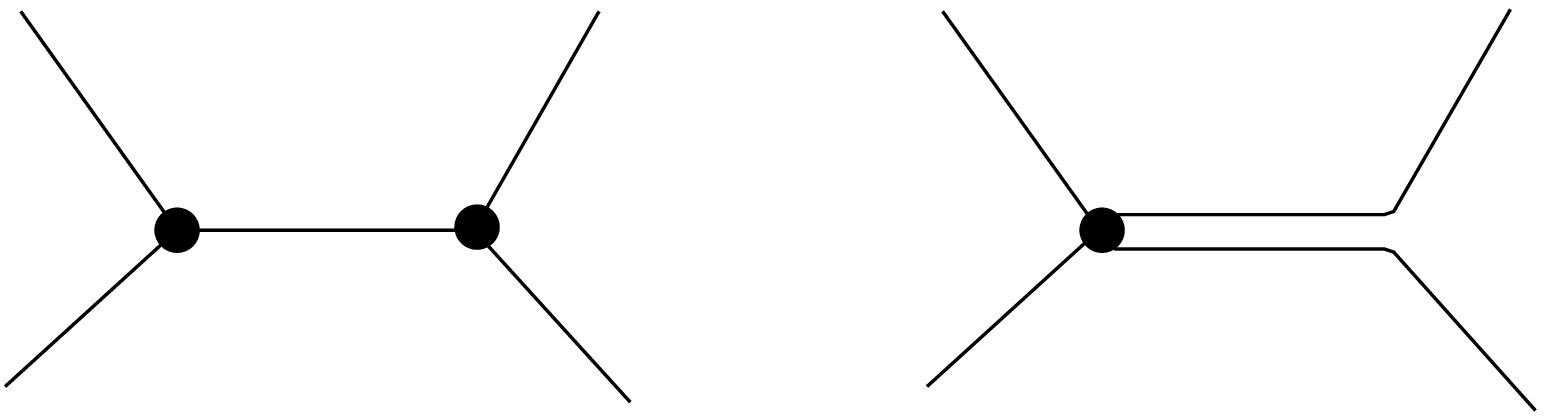}
$$

Similarly, to prove that $\Xi_{f_2,f_1}$ is unitary, we may suppose
$f_{1,2}$ are close to each other. We already know
$$\Xi_{f_2,f_1}\circ\Xi_{f_1,f_2}=\Xi_{f_2,f_2}=1;$$
checking that $\Xi_{f_2,f_1}$ is the adjoint of $\Xi_{f_1,f_2}$ is
straightforward.

Only few things remain now. Condition (2) holds evidently. We have to prove
$\eta(\overline\Si)=\overline{\eta(\Si)}$. Let $f:nD\rr\Si$ be a glueing and let
$r:D\rr\overline D$ be the reflection with respect to the vertical diameter.
Let $f':nD\rr\overline\Si$ be the composition
$nD\rTeXto^r n\overline D\rTeXto^{\overline f}\overline\Si$. The map
$*^{\otimes n}:\ha^{\otimes n}\rr\overline{\ha^{\otimes n}}$ sends
$\eta_f(\Si)$ onto $\overline{\eta_{f'}(\overline\Si)}$. It is easy to check
that it gives us a well-defined and functorial identification of
$\eta(\Si)$ with  $\overline{\eta(\overline\Si)}$.

Finally, we have to check that the quantum group coming from $\eta$
is indeed $\ha$. It is an easy exercise.

\section{The role of Drinfeld double}

Recall (the figure at the top of page 9) that $\eta(\Si)$ is a $\ha$-module for each black bead
and a $\ti\ha$-module for each white bead of $\Si$.
These $\ha$ and $\hat$ actions obviously commute with each other with
the exception of neighbouring beads, when the actions form Weyl
algebra (cf. \cite{nill}). If we take a boundary circle of $\Si$, all
actions combine together to an action of an associative algebra, which
is $\eta$ of the following figure (multiplication is given by glueing
the internal circle with the external circle of another copy of the
neckface on the figure; the action on $\eta(\Si)$ is given by glueing
the internal circle with the circle on $\Si$):
$$
\epsfxsize 25mm \epsfbox{weyl.epsi}
$$

When the boundary circle contains just one black and one white bead,
we will call it a {\em dyonic hole}. In this case the $\ha$ action and
the $\hat$ action combine to an action of the Drinfeld double $\do$,
i.e. to $\eta(A)$, where $A$ (together with nice glueing $D\du D\rr A$
giving rise to $\ha\otimes\hat\rr\do$) is
$$
\epsfxsize 20mm \epsfbox{doub.epsi}
$$
Thus $\eta(\Si)$ becomes a $\do$-module for every dyonic hole. $\ha$
and $\hat$ actions on $\eta(\Si)$ correspond to order and disorder
operators and at dyonic holes they form Drinfeld doubles.

Of course, the Drinfeld double is not just an associative algebra,
but a quasitriangular FQG; this will lead us to a braid group statistics
of the dyonic holes, given by the $R$-matrix of $\do$. The
comultiplication in $\do$ is drawn in the left part of the following
picture:
$$
\epsfxsize 80mm \epsfbox{Rmat.epsi}
$$
It presents a neckface $\Si$ together with a nice glueing $A\du
A\rr\Si$, giving $\do\otimes\do\rr\eta(\Si)$. If $\do$ acts on the
external circle of $\Si$, we have a map
$\do\otimes\eta(\Si)\rr\eta(\Si)$,
i.e. $\do\otimes(\do\otimes\do)\rr\do\otimes\do$. As in the case of the
comultiplication on $\ha$, it is given by 
$a\otimes(b\otimes c)\mapsto a_{(1)}b\otimes
a_{(2)}c$.

The right half of the picture presents another glueing $A\du A\rr\Si$;
it can be obtained by a map $\Si\rr\Si$ exchanging the interal
circles. This exchange correspods to the opposite comultiplication on
$\do$; if we have the element of $\eta(\Si)$ which is
$1\otimes1\in\do\otimes\do$ in the left decomposition of $\Si$, it
becomes the $R$-matrix $R\in\do\otimes\do$ in the right.

This exchange of two dyonic holes is the basic element from which the
braid group statistics of such holes is built up. It even proves that $\do$ is
quasitriangular.

\section{Conclusion}
Recall that our original aim was to find a lattice quantization of PL
T-duality. Clearly, it is still an open problem. The point is that we
have to admitt Hopf algebras that come from quantization of PL groups.
Finite dimension is (at least technically) good, so $\ha$ should be a
(restriced) quantum group at a root of 1. The most important difference
is that now $S^2\ne1$, as if turning $D$ for 360 degrees did not give
the identity.
The way out should be achieved by extending the category NEFA.

The functor $\eta$ is closely related to q-invariants of 3-folds
computed from the Drinfeld double. In fact, $\eta(\Si)$'s (for closed $\Si$'s)
are the state spaces of a TFT giving rise to these invariants
\cite{ker}. It would be
nice if it had topological applications for open $\Si$'s, too. It seems
reasonable that for $\Si$'s with dyonic holes only, it is connected with
ribbon tangles invariants.

There are good reasons to expect that the hypotetical vector spaces
$\eta(\Si)$ for the case of a root of 1 are closely related to the
state spaces of Chern-Simons theory, because 
1. in PL T-duality, WZNW action in the double gives the symplectic structure
on the phase space, and
2. in the finite quantum group case, the functor $\eta$ is closely related to
invariants of 3-folds computed from the double $\do(\ha)$.
Perhaps a part of the extension of NEFA should consist in  requiring
neckfaces to carry a complex structure.

\section*{Appendix: Finite quantum groups and the category $\FQ$}
The aim of this appendix is to fix the notation and to describe non-standard
definitions used in the text. We do not give standard definitions here
(see e.g. \cite{pres}).

A {\em finite quantum group} (FQG) $\ha$
is a finite-dimensional Hopf $C^*$-algebra.
We always rescale the inner product so that the multiplication map
$m:\ha\otimes\ha\rr\ha$ becomes projecting, i.e. $m|_{(\ker m)^\bot}$
is unitary.

The counit $\ep:\ha\rr\Co$ is clearly projecting, i.e. it is given by the
inner product with an $\ep\in\ha$, $\|\ep\|=1$ (usually, there is no confusion
between these two $\ep$'s, except for the definition of morphisms in the
category $\FQ$; we make a warning there). The dual counit $\ti\ep:\ha\rr\Co$
is given by the inner product with a positive multiple $\ti\ep$ of $1\in\ha$
such that $\ti\ep:\ha\rr\Co$ is projecting (or $\|\ti\ep\|=1$), i.e.
$\ti\ep=1/\|1\|$. Because
$\la\ep,1\ra=1$, we have $\|1\|\la\ep,\ti\ep\ra=1$.

The element $\ep\in\ha$ satisfies the identity
$$a\ep=\ep a=\la\ep,a\ra\ep$$
for any $a\in\ha$;
as a consequence,
$$\ep_{(1)}a\otimes\ep_{(2)}=\ep_{(1)}\otimes\ep_{(2)}S(a)$$
(both sides are equal to $\ep_{(1)}a_{(1)}\otimes\ep_{(2)}a_{(2)}S(a_{(3)})$).
We use the standard notation $a_{(1)}\otimes a_{(2)}$ for the coproduct of
$a$ and $a_{(1)}\otimes\dots\otimes a_{(n)}$ for the $n-1$-fold coproduct
of $a$.

The adjoint of the coproduct $\ha\rr\ha\otimes\ha$ gives  another associative
algebra structure $\ha\otimes\ha\rr\ha$. We rescale it by $\la\ep,\ti\ep\ra$
so that it becomes projecting. This map $\ti m$ is called
{\em  the dual multiplication}.

The dual FQG $\ti\ha$ is $\overline\ha=\ha^*$,
if we  exchange product with the dual product,
counit with the dual counit and $*$ with $S\circ *$
and take the complex
conjugate of the maps. This definition is clearly equivalent to 
the standard one, up to rescaling.

{\em Fourier transform} $\ha\rr\ti\ha$ is simply $*$ when $\overline\ha$
is understood as the FQG $\ti\ha$.

We organize FQG's into the {\em category $\FQ$}. A morphism $\ha_1\rr\ha_2$
is a linear map preserving the inner product (an isometric injection).
It has to preserve $S$ and $*$ and to preserve $m$, $\ti m$, $\ep$ and $\ti\ep$
{\em up to a positive factor}. {\em Warning: in this definition, $\ep$ and
$\ti\ep$ are understood as maps $\ha\rr\Co$ and not as elements of $\ha$.}
For example, if $N\ss G\ss H$ are finite groups and $N$ is normal in $G$,
there is a morphism between the group algebras $\Co G/N\rr\Co H$. It
is given by
$$gN\mapsto{|G|^{1\over2}\over|N| |H|^{1\over2}}\sum_{h\in gN}h.$$
Notice that for any $G$, $\|g\|^2=|G|$.

In FQG, the map $\ha\rr\ti\ha$ is a {\em covariant} automorphism.
We call it {\em duality}.

\subsection*{Acknowledgements}
I'd like to thank to Ctirad Klim\v c\'\i k for collaboration in the
early stage of this work and to Martin
Plech\v sm\'\i d
 for help with outwitting the computer. This work was
partially supported by the grant GA\v CR 201/96/0310.

\end{document}